\documentclass[global]{svjour}
\usepackage{graphicx}
\usepackage{amssymb, amsmath}
\usepackage{lscape}
\usepackage{color}
\newcommand{\fs}{\mbox{\ensuremath{.\!\!^{\rm s}}}}%  % fractional arcsecond symbol: 0.s0
\newcommand{\farcs}{\mbox{\ensuremath{.\!\!^{\prime\prime}}}}%  % fractional arcsecond symbol: 0.''0
\newcommand{\fdg}{\mbox{\ensuremath{.\!\!^\circ}}}%             % fractional degree symbol:     0.°0
\newcommand{\fm}{\mbox{\ensuremath{.\!\!^{\rm m}}}}%  % fractional magmitude symbol: 0.m0
\journalname{\emph{Astrophysics}}
\begin{document}
\title{ON THE NATURE OF UNCONFIRMED SUPERNOVAE}
\titlerunning{ON THE NATURE OF UNCONFIRMED SNe}
\author{L.~S.~Aramyan\inst{1}
\and A.~R.~Petrosian\inst{1}
\and A.~A.~Hakobyan\inst{1}
\and G.~A.~Mamon\inst{2}
\and D.~Kunth\inst{2}
\and M.~Turatto\inst{3}
\and V.~Zh.~Adibekyan\inst{4}
\and and T.~A.~Nazaryan\inst{1}
}                     % Do not remove
\institute{Byurakan Astrophysical Observatory, Armenia; \email{aramyan@bao.sci.am}
\and Institut d'Astrophysique de Paris, France
\and INAF - Osservatorio Astronomico di Padova, Italy
\and Centro de Astrof\'{i}sica da Universidade do Porto, Portugal}
\maketitle
\begin{abstract}
\begin{center}
\parbox{0.9\hsize}{\emph{We study the nature of 39 unconfirmed
supernovae (SNe) from the sky area covered by the
Sloan Digital Sky Survey (SDSS) Data Release 8 (DR8),
using available photometric and imaging data and intensive
literature search. We confirm that 21 objects are real SNe,
2 are Galactic stars, 4 are probable SNe, and
12 remain unconfirmed events. The probable types for 4 objects are suggested:
3 SNe are of probable type Ia, and SN~1953H is probable type II SN.
In addition, we identify the host
galaxy of SN~1976N and correct the offsets/coordinates of
SNe~1958E, 1972F, and 1976N.}}
\end{center}
\end{abstract}
\begin{center}
\parbox{0.9\hsize}{}
\end{center}
\section{Introduction}
\label{intro}

Because of the established role of the SNe to solve various problems
of modern astronomy, their continuous study is important.
In these studies, it is crucial to have homogeneous and well-defined
sample of different types of SNe and their host galaxies.
Most recently, paper \cite{hakobyan12} presented a new homogeneous data
set for 3876 SNe and their 3679 hosts. In this database, new observations were
used to reclassify and update the spectroscopic types of several SNe.
In addition, few objects turned out to be ``SN impostors'',
i.e., eruptions of Luminous Blue Variables (LBVs)
\cite{vanDyk05,smith11,pastorello12},
members of the class of Luminous Red Novae (LRN) \cite{pastorello12,kulkarni07},
or Galactic variable stars. In the database there are 39 unconfirmed SNe,
which are marked with the ``?'' symbol. Since these SNe are unconfirmed,
they are less useful for statistical investigations and
excluded from previous studies.

Currently the homogeneous data set for SNe and their host galaxies
presented in \cite{hakobyan12} is under scrutiny for several statistical studies.
To make the unconfirmed SNe useful for these studies,
it is important to clarify their nature. For this purpose,
we use available photometric and imaging data
and carry out intensive literature search.
The obtained results are presented below.

This paper is organized as follows: The sample selection and research method
are presented in Sects.~\ref{sample} and \ref{method}.
The results of study are given in Sect.~\ref{results} and summarized in Sect.~\ref{summary}.
Throughout this paper, we adopt the Hubble constant
$H_0=73 \,\rm km \,s^{-1} \,Mpc^{-1}$.

\section{The sample}
\label{sample}

The number of unconfirmed objects in the total sample of 3876 SNe
(discovered before March~22 of 2011 in the sky area covered by SDSS DR8)
is 39 \cite{hakobyan12}. The primary information of 39 question-marked SNe,
i.e., coordinates and/or offsets, and their host galaxies comes
from \cite{hakobyan12}. We used the Asiago Supernova Catalogue (ASC)
\cite{barbon99} to obtain other necessary information on the unconfirmed
SNe (magnitudes, epoch of discovery, discoverers, etc.).
In all cases the images of host galaxies of these SNe are available from the
First Palomar Observatory Sky Survey (POSS-I), POSS-II, and in a few cases also
from the UK Schmidt Telescope (UKST) plates of the southern hemisphere.
Since all the SN candidates are from SDSS coverage, we used the SDSS images as
a recent imaging data for them. Three SNe candidates (SNe~1991Y, 1992Y, and 2000af)
were discovered after 1990, and there is a possibility that additional observations for them
exist in the Multi-mission Archive at Space Telescope institute (MAST)
and/or European Southern Observatory (ESO) archive.
The search of these SNe in the mentioned archives shows no available images.

\section{The method of analysis}
\label{method}

In the current work to determine the nature and characterize the
unconfirmed SNe, special attention is paid to the primary information
(published mostly in the International Astronomical Union (IAU) circulars)
about the objects. Then we carefully examined the position of the SN on available
images of the host galaxies and conducted a deep literature search.
It is important to note that the SNe coordinates and offsets are reported
with different levels of accuracy, but fortunately, the precision is
generally within $1''$.
We assumed that the SN is real when the object is visible in
two or more images obtained within 400~days from the discovery date,
and is not visible in other images taken with more than 400~days from
the discovery. Special attention was paid to 7 objects
(SNe~1950O, 1951J, 1953H, 1954ad, 1955Q, 1955R, and 1955S)
with two POSS-I \emph{O} and \emph{E} plates available on the same day.
In such cases, there is a small
possibility that the visible object is a slow moving asteroid.
For all these cases astrometric comparison of
the positions of SN candidates was performed on available images.
The position of
the object in different images is the same,
within the $1\farcs7$ precision of the measurement.
Hence, misidentification of SN as an asteroid is unlikely,
considering apparent magnitudes and proper motion limit for an object at a distance
$\sim34$~AU due to reflex motion of the Earth \cite{ivezic01}.
We assumed that the object is not a SN and probably is a projected Galactic
star or a SN impostor when is visible in two or more images
taken more than 400~days apart.
In the cases where only one image for the candidate is available, its stellar
like nature was studied, and photometry was performed.
We counted the object as a probable SN if the image did not appear as a defect.
Finally, we suggested to keep question marks for SNe and count them as
unconfirmed SNe for the remaining cases, when even the original image of the candidate
was absent and no other imaging and useful information was available.

Determination of the probable SN type was performed according to the morphology of SNe hosts,
absolute discovery magnitudes and/or colors of SNe candidates,
and SNe position relative to morphological details,
such as spiral arms and HII regions. Absolute magnitudes at discovery of the objects
were calculated and compared with the mean magnitudes of different SNe types \cite{richardson02}.
To calculate the absolute magnitudes of SN candidates,
we used the method described in \cite{hakobyan12}. If the SN is discovered in
an early-type galaxy, we count it as a probable type Ia.
Type Ia was suggested also for objects located far from the disk structure
of the spiral host (in the case where the host galaxy is inclined).

It is well known that core-collapse (CC) SNe avoid early-type galaxies \cite{hakobyan08},
preferably being associated with disk structure \cite{hakobyan09},
and they are generally discovered in spiral arms (e.g., \cite{maza76,bartunov94})
or star-forming regions (e.g., \cite{vandyk92,vandyk99,anderson08,anderson12}),
but because type Ia SNe also show a correlation with spiral arms \cite{bartunov94}
(though not as strong as CC), and there is a chance of projection of
type Ia SN on HII regions, the association to spiral arms or HII regions
was not used for the classification.
Confirmed or probable SNe are classified as CC according to their
blue colors and/or absolute magnitudes only.

\section{Results}
\label{results}

\subsection{Confirmed SNe}
\label{results1}

According to the aforementioned criteria, 21 unconfirmed SNe out of 39 are shown
to be real SN events. These 21 objects are collected in Table~\ref{confSNe},
with the names and morphology of their hosts, SNe coordinates, offsets,
and magnitudes (photometric band indicated) upon discovery from IAU circulars, as well as with their probable types.
A magnitude without band means that the observation has not been made in a standard photometric system
(e.g., those reported in the discovery announcement as photographic, blue plate, red plate, CCD without filter, and so on).
Anonymous galaxies are listed with the letter ``A'' followed by the coordinates.

\begin{table}[t]
\begin{center}
\caption{The list of confirmed objects.}
\label{confSNe}
\tabcolsep 1.3pt
\begin{tabular}{lllllllll}
\hline
\hline
\multicolumn{1}{c}{SN} & \multicolumn{1}{c}{Galaxy} & Morph. & \multicolumn{1}{c}{$\alpha_{\rm SN}$} & \multicolumn{1}{c}{$\delta_{\rm SN}$} & E/W offset & N/S offset & Discovery \emph{mag} & Probable type \\
\hline
1950O & A161509+1857 & Sbc & 16 15 09.3 & +18 57 14.4 & \,\,\,\,\, 10.7E & \,\,\,\,\, 10.6S & \,\,\,\,\,\,\,\,\, B17 & \\
1951J & MCG+00--15--01 & SBc & 05 37 52.6 & +00 07 36.9 & \,\,\,\,\, 24W & \,\,\,\,\, 13N & \,\,\,\,\,\,\,\,\, B17.5 & \\
1953H & A110342+4951 & S & 11 03 38.7 & +49 50 29.4 & \,\,\,\,\, 1E & \,\,\,\,\, 3S & \,\,\,\,\,\,\,\,\, B17 & \,\,\,\,\,\,\,\,\, II:\\
1954ad & UGC4467 & Sb & 08 32 46.3 & +00 13 33.7 & \,\,\,\,\, 7.2W & \,\,\,\,\, 5.9S & \,\,\,\,\,\,\,\,\, B17.5 & \\
1955Q & A105606+2409 & Sm & 10 56 08.4 & +24 09 19 & \,\,\,\,\, 4W & \,\,\,\,\, 7S & \,\,\,\,\,\,\,\,\, B17.5 & \\
1955R & UGC7740 & Sc & 12 34 42.4 & +49 19 54.8 & \,\,\,\,\, 3.8E & \,\,\,\,\, 26N & \,\,\,\,\,\,\,\,\, B18 & \\
1955S & UGC9933 & SBab & 15 36 40.6 & +43 32 38 & \,\,\,\,\, 16.2W & \,\,\,\,\, 15.8N & \,\,\,\,\,\,\,\,\, B17.5 & \\
1968J & PGC50284 & S0 & 14 05 52.1 & +53 07 32.3 & \,\,\,\,\, 2W & \,\,\,\,\, 12S & \,\,\,\,\,\,\,\,\, \,\,\,\,16 & \,\,\,\,\,\,\,\,\, Ia\\
1969G & A123342+0553 & Sb & 12 33 50 & +05 53 42.8 & \,\,\,\,\, 4E & \,\,\,\,\, 7N &  & \\
1970M & A104818+1403 & S & 10 48 12 & +14 03 15.5 & \,\,\,\,\, 13E & \,\,\,\,\, 1N & \,\,\,\,\,\,\,\,\, \,\,\,\,16.5 & \\
1972F & MCG+09--20--97 & Sa & 12 07 11.1 & +53 40 32.2 & \,\,\,\,\, 17E & \,\,\,\,\, 15N & \,\,\,\,\,\,\,\,\, \,\,\,\,16 & \\
1974D & NGC3916 & Sbc & 11 50 55 & +55 09 07.9 & \,\,\,\,\, 34E & \,\,\,\,\, 31N & \,\,\,\,\,\,\,\,\, \,\,\,\,15.5 & \\
1976A & NGC5004A & SBb & 13 11 01.1 & +29 34 59.9 & \,\,\,\,\, 7W & \,\,\,\,\, 18N & \,\,\,\,\,\,\,\,\, \,\,\,\,16.5 & \\
1976N & A073200+6513 & S & 07 31 51.4 & +65 12 38.1 & \,\,\,\,\, 7E & \,\,\,\,\, 6S & \,\,\,\,\,\,\,\,\, \,\,\,\,15 & \\
1980A & MCG+05--29--64A & SBb & 12 20 28.3 & +31 10 10.5 & \,\,\,\,\, 9E & \,\,\,\,\, 9S & \,\,\,\,\,\,\,\,\, \,\,\,\,15.5 & \\
1980B & MCG+09--19--42 & SBc & 11 19 54.3 & +54 27 46.2 & \,\,\,\,\, 16E & \,\,\,\,\, 0N & \,\,\,\,\,\,\,\,\, \,\,\,\,16 & \\
1980C & A134524+4745 &  &  &  & \,\,\,\,\, 7W & \,\,\,\,\, 0N & \,\,\,\,\,\,\,\,\, \,\,\,\,17.5 & \\
1980E & A131942+3414 & Sm & 13 19 42.5 & +34 14 03.9 & \,\,\,\,\, 8W & \,\,\,\,\, 7S & \,\,\,\,\,\,\,\,\, \,\,\,\,16 & \\
1982X & UGC4778 & Sc & 09 07 01.4 & +50 43 08.7 & \,\,\,\,\, 92W & \,\,\,\,\, 23N & \,\,\,\,\,\,\,\,\, V16.5 & \\
1982Y & UGC5449 & SBc & 10 08 00.5 & +68 21 58.7 & \,\,\,\,\, 12E & \,\,\,\,\, 5N & \,\,\,\,\,\,\,\,\, V17 & \\
2000af & A114855--0058 &  & 11 48 55.7 & \,--00 58 36.1 & \,\,\,\,\, 0E & \,\,\,\,\, 0N & \,\,\,\,\,\,\,\,\, R19.8 & \\
\hline
\\
\end{tabular}
\end{center}
\end{table}

\textbf{\emph{SN~1950O.}} SN~1950O as a 17 magnitude object was discovered on
POSS-I \emph{O} plate \cite{mueller92}. This object is visible in both POSS-I \emph{O}
and \emph{E} images (both taken on April~17, 1950).
No star-like object is visible in SN position on POSS-II \emph{J} and
\emph{F} plates, as well as on the SDSS image.

\textbf{\emph{SN~1951J.}} SN~1951J as a 17.5 magnitude object was discovered on POSS-I \emph{O}
plate \cite{mueller90}. The SN is visible on POSS-I \emph{O} and \emph{E} images (both on
November~29, 1951) and it is not visible on the Science and Engineering Research
Council (SERC) southern sky survey \emph{J}, \emph{ER} and \emph{I},
POSS-II \emph{J}, \emph{F}, and \emph{N} plates,
as well as on the SDSS image.

\begin{figure}[t]
\begin{center}$
\begin{array}{cc}
\includegraphics[width=0.4\hsize]{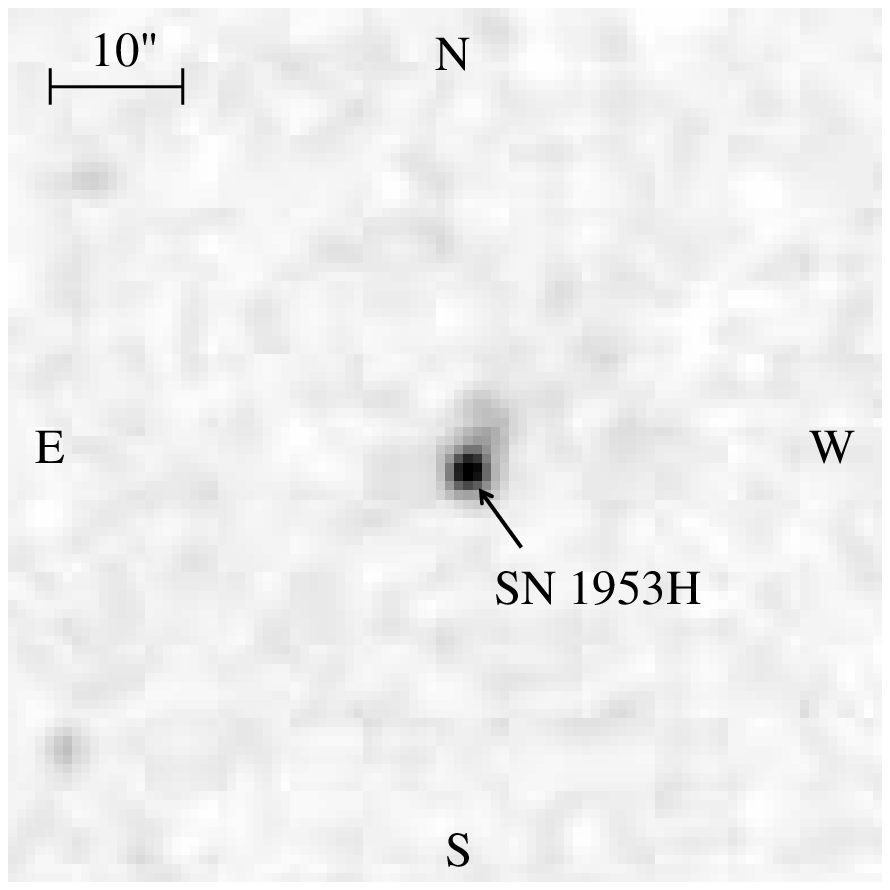} &
\includegraphics[width=0.4\hsize]{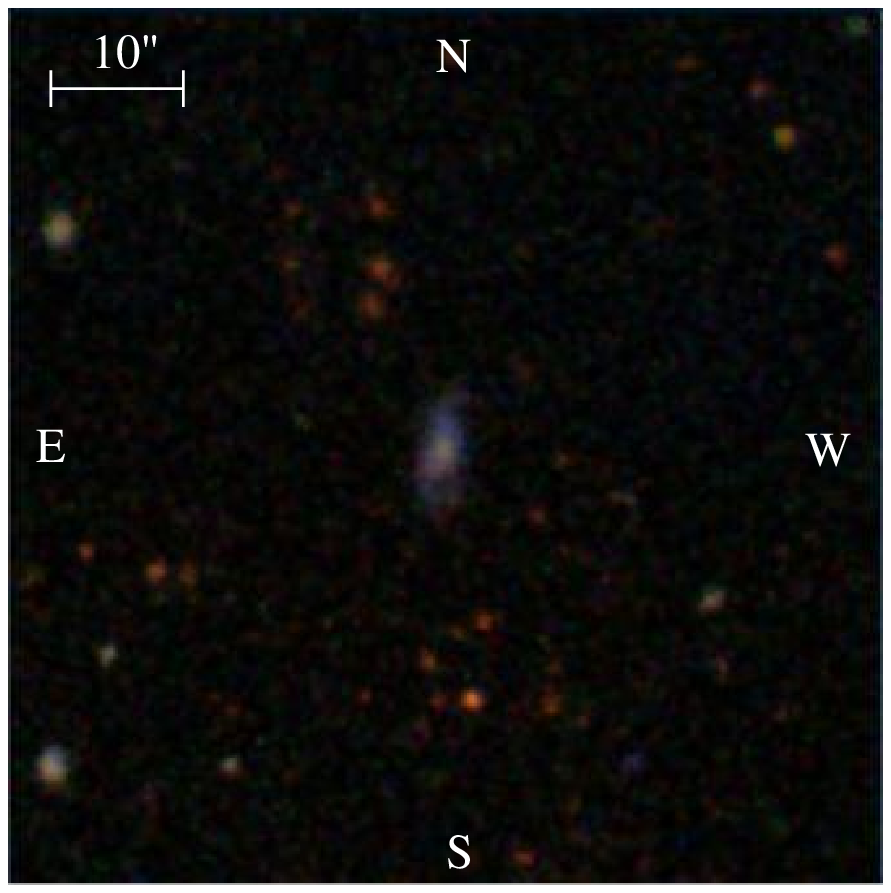}
\end{array}$
\end{center}
\caption{Left: the SN~1953H in the POSS-I \emph{O} image.
The SN name and position (marked by the arrow)
are shown. Right: the host galaxy of SN~1953H in the SDSS image.
In all images, north is up and east is to the left.}
\label{1953H}
\end{figure}

\textbf{\emph{SN~1953H.}} SN~1953H was discovered by \emph{Notni \& Oleak},
but no appropriate publication with finding chart is available.
The SN is visible on both POSS-I \emph{O} (Fig.~\ref{1953H}, left)
and \emph{E} plates (both on March~17, 1953) and is not visible in
all POSS-II and SDSS images (e.g., Fig.~\ref{1953H}, right).
Since the color of the SN is $B-R=-0.3$, we conclude that it was discovered
very close to explosion and probably is a type II.

\textbf{\emph{SN~1954ad.}} SN~1954ad as a 17.5 magnitude object was discovered
on POSS-I plates \cite{mueller91}. It is well seen on both POSS-I \emph{O} and \emph{E} images
(epoch: December~21, 1954) and absent on all plates of SERC, POSS-II, as well as
on the SDSS DR8 images.

\textbf{\emph{SN~1955Q.}} SN~1955Q is reported in \cite{pollas92} following inspection of
POSS-I plates. The object is visible on both POSS-I \emph{O} and \emph{E} images
(epoch: March~25, 1955) and not visible on plates of POSS-II and SDSS images.

\textbf{\emph{SN~1955R.}} SN~1955R as a 18 magnitude object was discovered on POSS-I plates \cite{mueller+92}.
The object is visible on both POSS-I \emph{O} and \emph{E} images (epoch: April~15, 1955)
and not visible on all POSS-II plates and SDSS images.

\textbf{\emph{SN~1955S.}} SN~1955S as a 17.5 magnitude object was discovered on POSS-I plates \cite{mueller93}.
The object is visible on both POSS-I \emph{O} and \emph{E} images (epoch: April~1, 1955)
and not visible on all POSS-II plates and SDSS images.

\textbf{\emph{SN~1969G.}} SN~1969G was reported as an $m_{\rm pg}=18^{\rm m}$ object in \cite{kowal70}.
In this paper, a low-quality print for SN~1969G is presented too.
The SN was visible in 3 separate plates, taken between April~13 and June~16, 1969 \cite{kowal70}.
No object is visible on the calculated position of SN~1969G on
POSS-I \emph{O} and \emph{E}, POSS-II \emph{J}, \emph{F}, and \emph{N},
as well as on the SDSS images. Since the object was visible in 3 different images \cite{kowal70} and not visible
in all other available images, we suggest that this was a real SN.

\textbf{\emph{SN~2000af.}} SN~2000af was discovered by \emph{Schaefer} \cite{schaefer00}. Later, a spectral observation of
this object is reported \cite{filippenko00}. According to \cite{filippenko00}, the spectrum of SN~2000af, taken in poor seeing,
shows no obvious supernova features. No star-like source is visible at the SN position on
all plates of POSS-I, POSS-II, SERC, and SDSS images.
Since the SN~2000af was visible in 4 different images \cite{schaefer00}, we confirm it as real SN event.

For 11 objects (SNe~1968J, 1970M, 1972F, 1974D, 1976A, 1976N, 1980A, 1980B, 1980C, 1982X, and 1982Y)
discovered at the Konkoly Observatory, two different plates with SN event are available \cite{tsvetkova08}.
In addition, four such plates are available for SN~1980E \cite{tsvetkova08}.
In analogy to the other objects discussed above, we suggest to consider all these SNe as real events.

The first announcement of SN~1972F \cite{lovas72} reports that the object was $34''$ east and $14''$ north
from the nucleus of the host galaxy. SN~1972F is included in \cite{barbon99} with offsets $17''$ east and $15''$ south.
Using imaging information in \cite{tsvetkova08}, we corrected and recalculated the offsets of SN~1972F
to $17''$ east and $15''$ north.

\begin{figure}[t]
\begin{center}$
\begin{array}{cc}
\includegraphics[width=0.4\hsize]{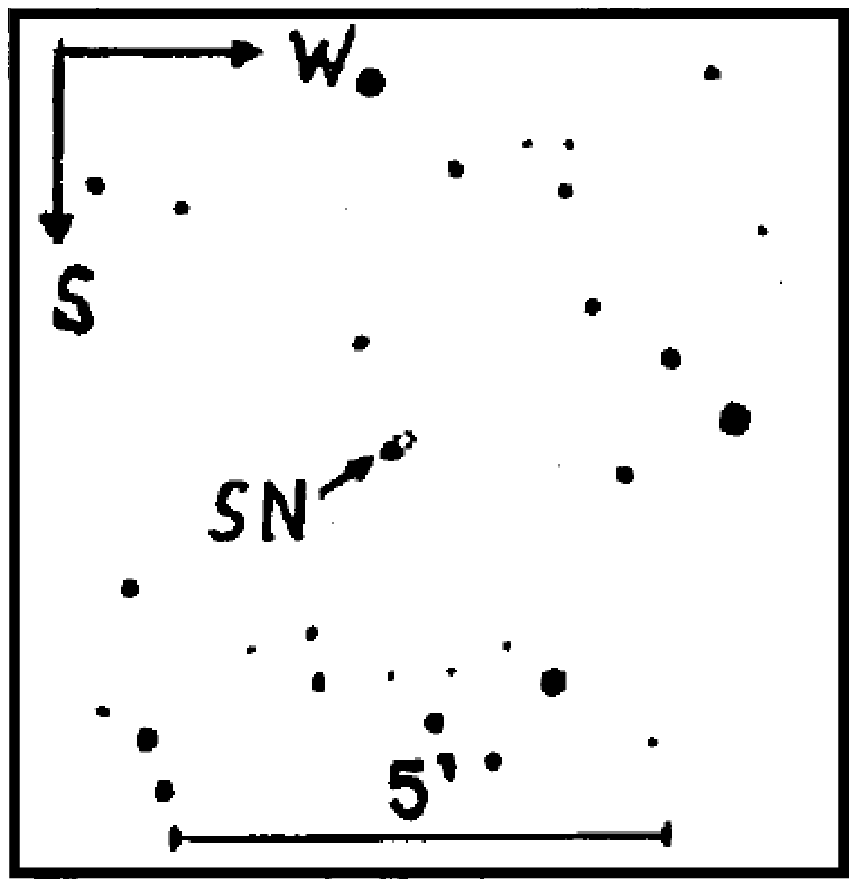} &
\includegraphics[width=0.4\hsize]{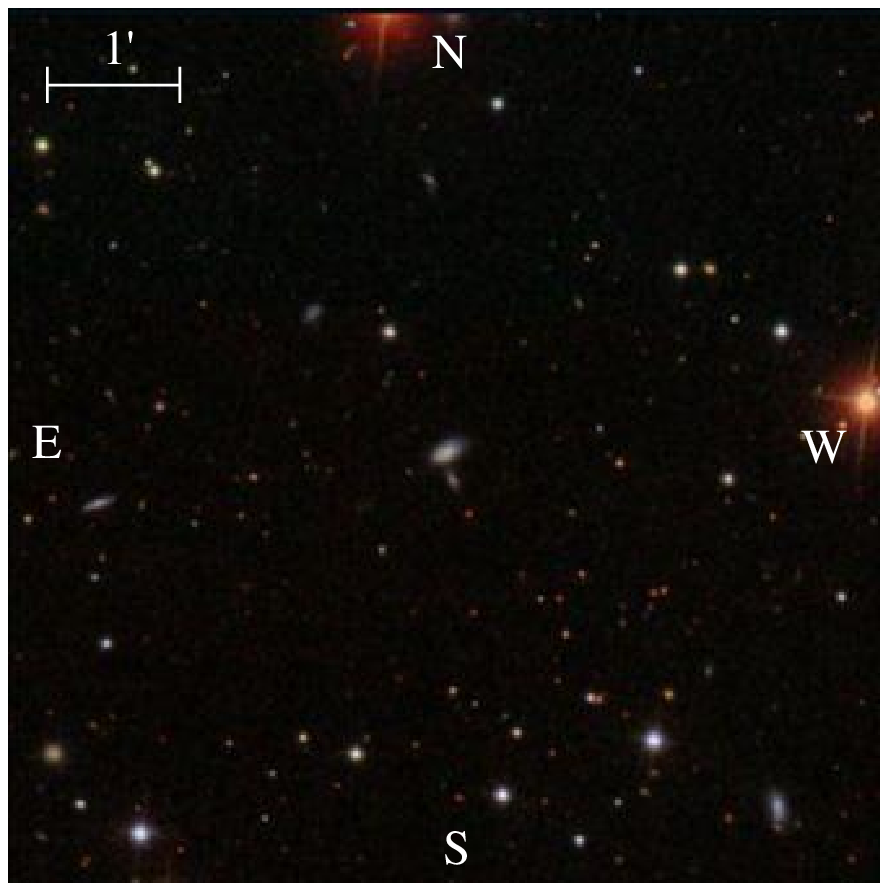}
\end{array}$
\end{center}
\caption{Left: the identification chart for SN~1976N,
which is available from \cite{lovas79}.
The SN position (marked by the arrow) is shown.
Right: the SDSS image of the same field.
In all images, north is up and east is to the left.}
\label{1976N}
\end{figure}

Finally, the identification chart (Fig.~\ref{1976N}, left) is available for SN~1976N \cite{lovas79}.
Comparing the identification chart with the SDSS image of the field, we identified
the host galaxy of the SN (Fig.~\ref{1976N}, right). The host is a spiral galaxy with small angular size
(about $15''$). The SDSS identification of the host is J073150.33$+$651244.1,
and the calculated coordinates of the possible SN
(via the host nucleus coordinates and SN offsets) are
${\rm \alpha = 07^{h}31^{m}51\fs44}$,
${\rm \delta = +65^\circ12'38\farcs1}$ (J2000.0).

\subsection{Possible Galactic stars}
\label{results2}

Out of the 39 unconfirmed SNe two turned out to be, with high probability, Galactic stars or SN impostors. These two objects are listed in Table~\ref{SNimps}
where the names and morphology of the hosts, objects coordinates, their offsets,
and magnitudes are given too.

\begin{table}[h]
\begin{center}
\caption{The list of not real SN explosions.}
\label{SNimps}
\tabcolsep 3pt
\begin{tabular}{llllllll}
\hline
\hline
\multicolumn{1}{c}{SN} & \multicolumn{1}{c}{Galaxy} & Morph. & \multicolumn{1}{c}{$\alpha_{\rm SN}$} & \multicolumn{1}{c}{$\delta_{\rm SN}$} & E/W offset & N/S offset & Discovery \emph{mag} \\
\hline
1954Y & MCG+03--35--37 & SBa & 13 54 30.2 & +15 02 38.7 & \,\,\,\,\, 13W & \,\,\,\,\, 0N & \,\,\,\,\,\,\,\,\, V19.3 \\
1973O & NGC7337 & SBb & 22 37 24.5 & +34 21 59.4 & \,\,\,\,\, 26W & \,\,\,\,\, 28S & \,\,\,\,\,\,\,\,\, V19.0 \\
\hline
\end{tabular}
\end{center}
\end{table}

Below are descriptions of these cases.

\begin{figure}[t]
\begin{center}$
\begin{array}{ccc}
\includegraphics[width=0.3\hsize]{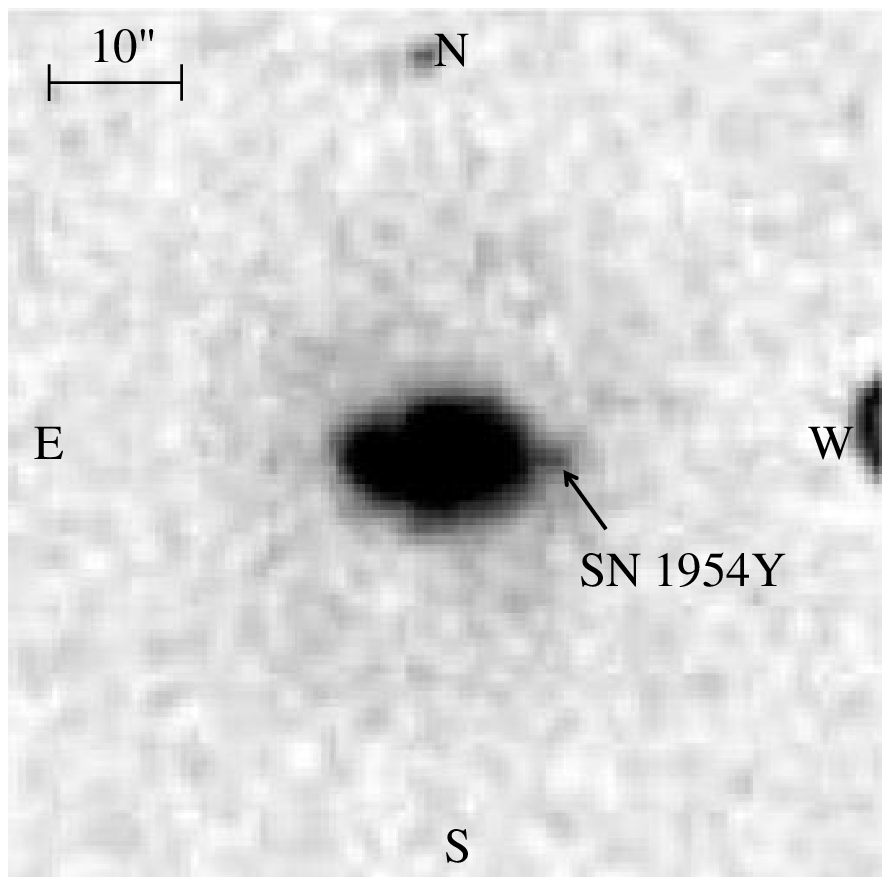} &
\includegraphics[width=0.3\hsize]{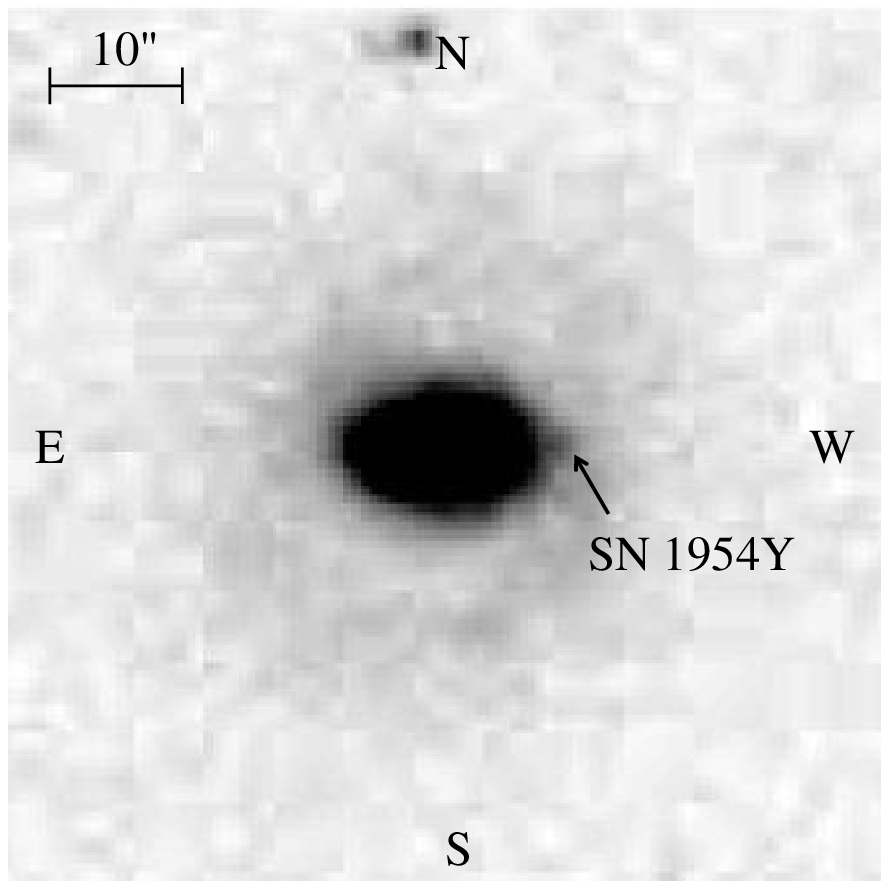} &
\includegraphics[width=0.3\hsize]{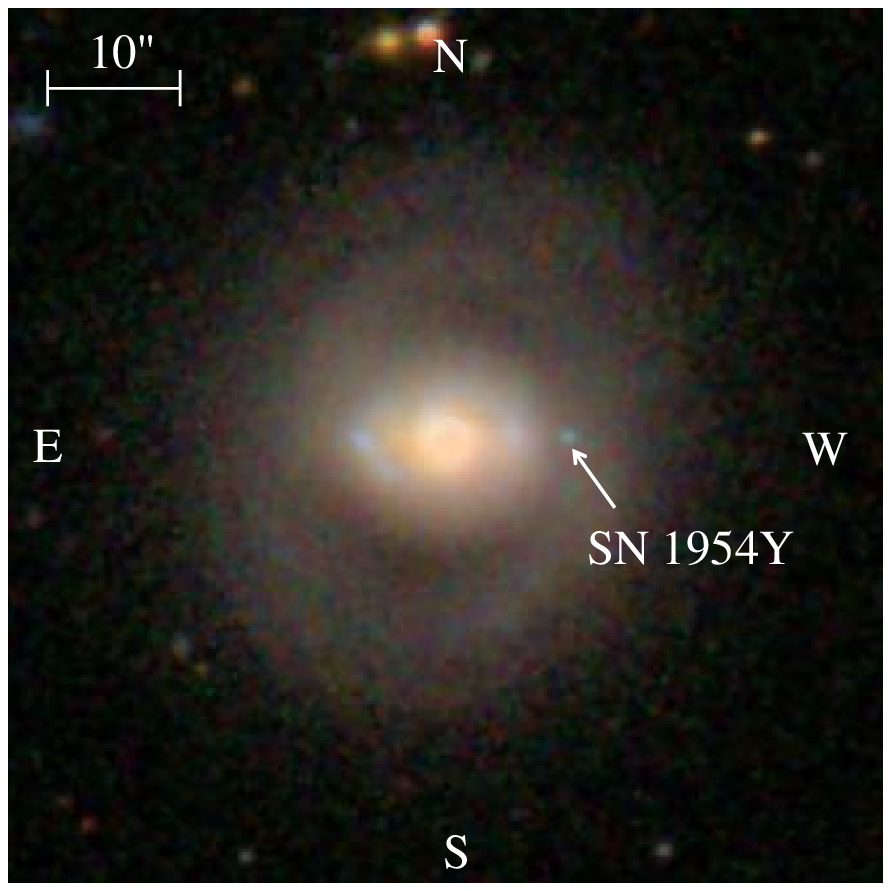}
\end{array}$
\end{center}
\caption{Left: POSS-I \emph{O} image of the SN~1954Y.
Middle: SN~1954Y in POSS-II \emph{J} image.
Right: SN~1954Y in SDSS image.
The SN position (marked by the arrow) is shown.
In all images, north is up and east is to the left.}
\label{1954Y}
\end{figure}

\textbf{\emph{SN~1954Y.}} The first publication on SN~1954Y is \cite{kowal74}.
SN~1954Y was discovered on POSS-I \emph{O} plate at $V=19.3$ magnitude.
Figure~\ref{1954Y} (left) presents the POSS-I \emph{O} (epoch: April~1, 1954) discovery image of the SN.
On the position of SN~1954Y a star-like source is visible.
The same object is also visible on POSS-II \emph{J} (at magnitude of 19.1 on May~19, 1993, Fig.~\ref{1954Y}, middle)
and \emph{F} (epoch: April~21, 1996) images.
In the SDSS DR8 image (epoch: May~13, 2005, Fig.~\ref{1954Y}, right),
on the position of the SN a star-like object with $g=18.23$ magnitude is visible.
The object is not visible in POSS-I \emph{E} (epoch: April~1, 1954) and POSS-II \emph{N} (epoch: March~10, 1997) images.
Taking into account that a star-like object is visible in 4 different images,
with very different epochs, we conclude that the candidate is not a SN, but rather a SN impostor.

\begin{figure}[t]
\begin{center}$
\begin{array}{cc}
\includegraphics[width=0.4\hsize]{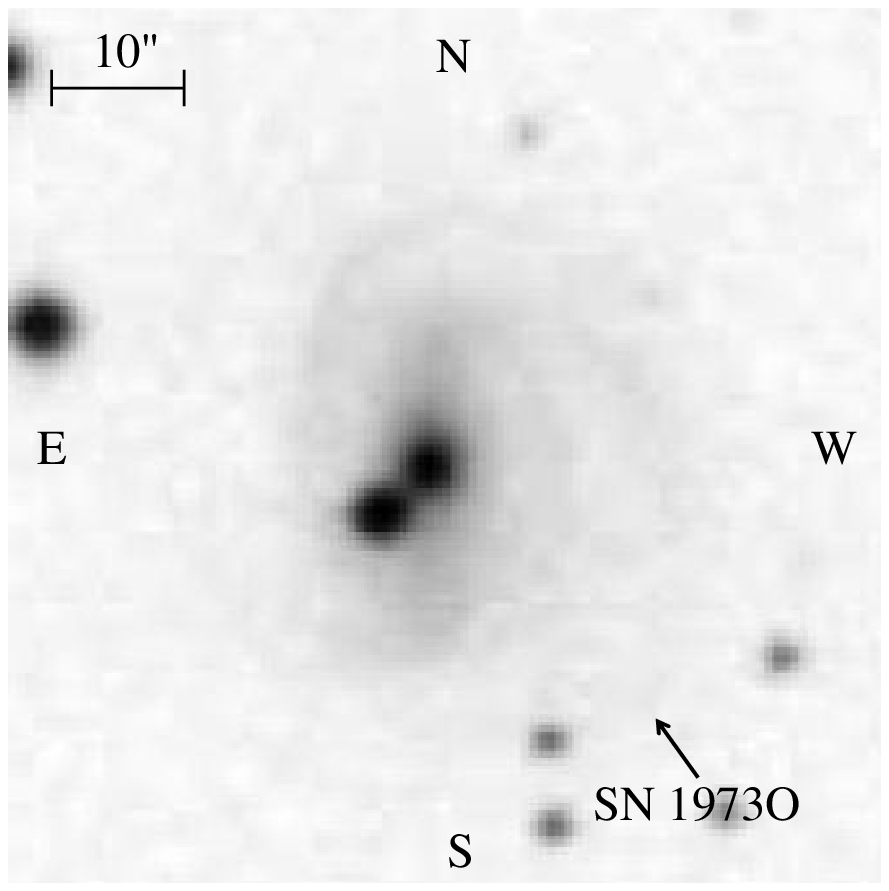} &
\includegraphics[width=0.4\hsize]{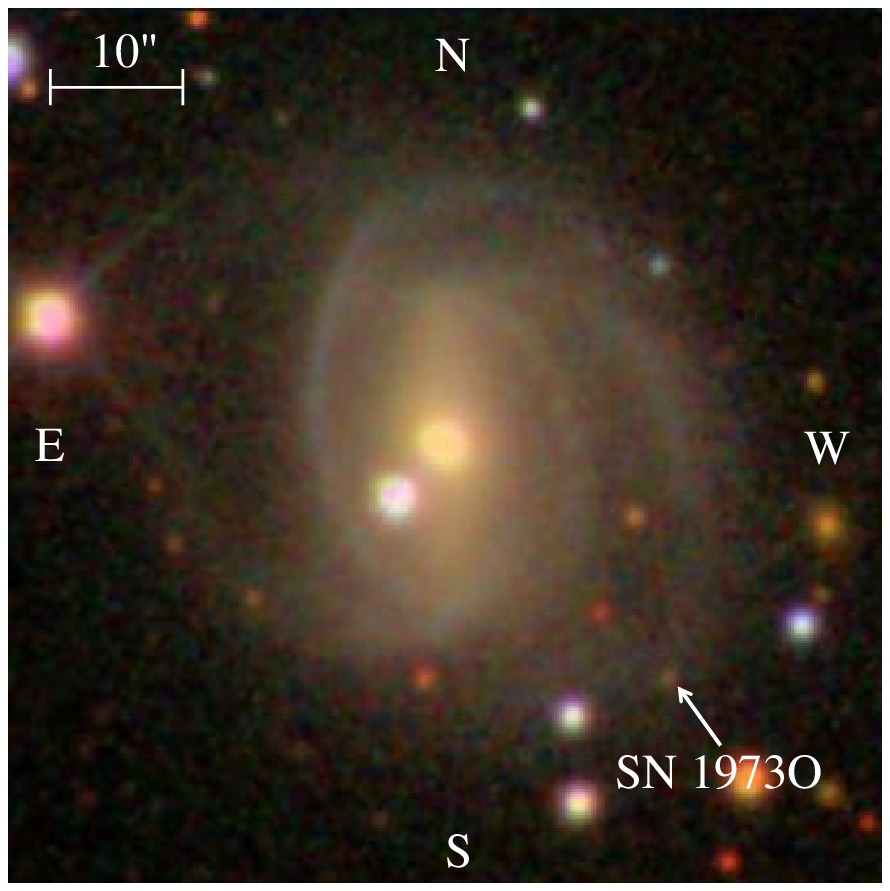}
\end{array}$
\end{center}
\caption{Left: the position of SN~1973O in POSS-II \emph{J} plate.
Right: the SN~1973O in the SDSS image.
The SN position (marked by the arrow) is shown.
In all images, north is up and east is to the left.}
\label{1973O}
\end{figure}

\textbf{\emph{SN~1973O.}} SN~1973O is reported in \cite{kormendy73} with 19.0 photo-visual magnitude.
It was located $26''$ west and $28''$ south of the galaxy nucleus.
No star-like object is visible at the position calculated for the SN on
POSS-I \emph{O}, \emph{E}, and POSS-II \emph{J}, \emph{F}, and \emph{N} plates (e.g., Fig.~\ref{1973O}, left),
but a stellar object with $g=22.33$ magnitude is clearly visible in the SDSS DR8 image
(Fig.~\ref{1973O}, right). We have estimated the magnitude from all the available material
(Table~\ref{SN1973O}) and conclude that the object varied in a non-monotonic way over a time interval of 53 years.
Considering that the candidate of SN is visible on the SDSS image, we conclude that SN~1973O
is a foreground Galactic variable star.

\begin{table}[h]
\begin{center}
\caption{The brightness of SN~1973O.}
\label{SN1973O}
\tabcolsep 3pt
\begin{tabular}{lll}
\hline
\hline
Epoch & \emph{mag} & Source \\
\hline
August 12, 1953 & $B>21$ & POSS-I \emph{O} image \\
September 4, 1973 & $V=19.0$ & \cite{kormendy73} \\
September 8, 1989 & $B>22.5$ & POSS-II \emph{J} image \\
May 30, 2006 & $g=22.3$ & SDSS DR8 image \\
\hline
\end{tabular}
\end{center}
\end{table}

\subsection{Probable SNe}
\label{results3}

Four objects are suggested to be probable SNe out of the 39 unconfirmed SNe.
Table~\ref{probSNe} presents information on them and
their host galaxies. The last column of Table~\ref{probSNe} gives the probable types
of two of them.

\begin{table}[h]
\begin{center}
\caption{The list of probable SNe.}
\label{probSNe}
\tabcolsep 1.3pt
\begin{tabular}{lllllllll}
\hline
\hline
\multicolumn{1}{c}{SN} & \multicolumn{1}{c}{Galaxy} & Morph. & \multicolumn{1}{c}{$\alpha_{\rm SN}$} & \multicolumn{1}{c}{$\delta_{\rm SN}$} & E/W offset & N/S offset & Discovery \emph{mag} & Probable type \\
\hline
1958E & MCG+07--07--72 & SBa & 03 20 40.2 & +42 48 10.2 & \,\,\,\,\, 12.2W & \,\,\,\,\, 4.4S & \,\,\,\,\,\,\,\,\, \,\,\,\,17.5 &  \\
1986P & NGC5763 & S0 & 14 48 58.1 & +12 29 18.3 & \,\,\,\,\, 8W & \,\,\,\,\, 6S & \,\,\,\,\,\,\,\,\, B17 & \,\,\,\,\,\,\,\,\, Ia \\
1991Y & A171436+5719 & S0 & 17 14 37.6 & +57 18 26.4 & \,\,\,\,\, 0E & \,\,\,\,\, 7N & \,\,\,\,\,\,\,\,\, B19 & \,\,\,\,\,\,\,\,\, Ia \\
1992Y & NGC3527 & SBa & 11 07 19.1 & +28 31 33.8 & \,\,\,\,\, 13E & \,\,\,\,\, 6S & \,\,\,\,\,\,\,\,\, B18.5 &  \\
\hline
\end{tabular}
\end{center}
\end{table}

\textbf{\emph{SN~1958E.}} SN~1958E was discovered in February~10, 1958 \cite{humason60} at magnitude $17\fm5$.
The object was not visible on POSS-I \emph{O} and \emph{E},
as well as on all POSS-II plates and SDSS images. Since SN~1958E was not detected on two POSS-I plates,
taken 52~days before the discovery, and considering that the limiting magnitude of POSS-I
plates reaches to $21^{\rm m}$ \cite{abell59}, we conclude that SN~1958E did not explode at that time.
In any case, to be sure that the image of SN is not a photographic defect on
Palomar 48~inch telescope plate, we studied the SN profile and estimate the SN magnitude on
the SDSS $g$-band image using the magnitudes of nearby stars.
We found that it cannot be a photographic defect, the profile of SN is star-like,
and its magnitude is about $17^{\rm m}$, close to the value reported in \cite{humason60}.
Because of the absence of more confirmation image for SN~1958E, we consider it as a probable SN.

No astrometric data for SN~1958E is presented in \cite{humason60}, while the ASC reports an offset of $11''$ west.
By means of an astrometric comparative study of \cite{humason60} and the SDSS images,
we now report new correct coordinates for the object,
${\rm \alpha = 50\fdg16782}$,
${\rm \delta = 42\fdg80285}$ (J2000.0)
and offsets from the host galaxy nucleus, $4\farcs4$ south and $12\farcs2$ west.

\textbf{\emph{SN~1986P.}} SN~1986P was first reported in \cite{mueller96} at
$17^{\rm m}$ on \emph{IIIa-J} plate,
taken in the course of a POSS-II survey with the 1.2-m Oschin Schmidt Telescope on July~2, 1986.
In \cite{mueller96} it is stressed that the SN cannot be a plate defect or an asteroid.
No star-like source is visible at the SN position on POSS-I, POSS-II \emph{F} (epoch: May~14, 1993),
and \emph{N} (epoch: March~12, 1997) plates, as well as on SDSS images.
Since the well-studied image of the SN is visible only on POSS-II \emph{J} plate \cite{mueller96}
but not in other available images, we suggest it to be a probable SN.
According to \cite{hakobyan12} the host of SN~1986P is an S0 type galaxy,
which means that this object can be type Ia SN.

\textbf{\emph{SN~1991Y.}} SN~1991Y as a $19^{\rm m}$ object was discovered
on POSS-II \emph{J} plate and reported by \cite{mueller+91}.
No object is visible at the SN position on POSS-I, POSS-II \emph{F} (epoch: June~30, 1992),
and \emph{N} (epoch: June~4, 1991) plates and SDSS images. The fact that the object is not visible on
POSS-II \emph{N} image, taken about 1.5~months after the \emph{J} plate,
could be easily explained by the shallower limiting magnitude $19\fm5$ \cite{reid91}
of POSS-II \emph{N} plates. According to \cite{hakobyan12} the host of SN~1991Y is an S0 type galaxy,
which means that this object can be type Ia SN.

\textbf{\emph{SN~1992Y.}} SN~1992Y as an $18\fm5-19^{\rm m}$ object was discovered on
POSS-II \emph{J} plate \cite{mueller++92} (epoch: May~2, 1992).
No star-like source is visible at the SN position on all POSS-I, POSS-II \emph{F} (epoch: May~21, 1998),
\emph{N} (epoch: January~13, 2000), and SDSS images.

\subsection{Unconfirmed SNe}
\label{results4}

Because of the lack of information, the nature of 12 objects out of 39 unconfirmed SNe
was not clarified and remain unconfirmed.

Table~\ref{uncoSNe} presents the list of 12 unconfirmed SNe, with data of these objects
and their host galaxies.
In all 12 cases, no star-like source is visible in POSS-I, POSS-II, and SDSS images.

\begin{table}[h]
\begin{center}
\caption{The list of unconfirmed SNe.}
\label{uncoSNe}
\tabcolsep 3pt
\begin{tabular}{llllllll}
\hline
\hline
\multicolumn{1}{c}{SN} & \multicolumn{1}{c}{Galaxy} & Morph. & \multicolumn{1}{c}{$\alpha_{\rm SN}$} & \multicolumn{1}{c}{$\delta_{\rm SN}$} & E/W offset & N/S offset & Discovery \emph{mag} \\
\hline
1950M & NGC3266 & SB0 & 10 33 17.9 & +64 45 00.6 & \,\,\,\,\, 2E & \,\,\,\,\, 3N & \,\,\,\,\,\,\,\,\, \,\,\,\,14.5 \\
1951I & NGC6181 & SBc &  &  &  &  & \,\,\,\,\,\,\,\,\, \,\,\,\,15.7 \\
1955C & NGC23 & SBb & 00 09 54.1 & +25 55 35.6 & \,\,\,\,\, 10E & \,\,\,\,\, 10N & \,\,\,\,\,\,\,\,\, V16 \\
1965O & A120224+4955 & S0 & 12 02 19.7 & +49 56 11.9 & \,\,\,\,\, 13W & \,\,\,\,\, 17N & \,\,\,\,\,\,\,\,\, \,\,\,\,17.5 \\
1966O & PGC34376 & E/S0 & 11 16 13 & +29 21 37 & \,\,\,\,\, 7.1W & \,\,\,\,\, 89.8S & \,\,\,\,\,\,\,\,\, \,\,\,\,15.5 \\
1968U & NGC4183 & Sc & 12 13 15 & +43 43 28.3 & \,\,\,\,\, 20W & \,\,\,\,\, 95N & \,\,\,\,\,\,\,\,\, \,\,\,\,14.5 \\
1969Q & NGC4472 & S0 & 12 29 54.8 & +08 00 01.6 & \,\,\,\,\, 120E & \,\,\,\,\, 0N & \,\,\,\,\,\,\,\,\, \,\,\,\,14 \\
1972T & MCG+05--32--01 & SBc & 13 20 21.2 & +31 30 54 & \,\,\,\,\, 4W & \,\,\,\,\, 0N & \,\,\,\,\,\,\,\,\, \,\,\,\,14 \\
1974A & NGC4156 & SBb & 12 10 48.3 & +39 28 37 & \,\,\,\,\, 15W & \,\,\,\,\, 15N & \,\,\,\,\,\,\,\,\, \,\,\,\,20 \\
1984U & NGC4246 & Scd & 12 18 00.3 & +07 10 55.2 & \,\,\,\,\, 33E & \,\,\,\,\, 14S & \,\,\,\,\,\,\,\,\, \,\,\,\,18 \\
1985K & A125916+2759 & S & 12 59 16.1 & +27 59 03.7 & \,\,\,\,\, 0.5W & \,\,\,\,\, 7S & \,\,\,\,\,\,\,\,\, V17.7 \\
1987R & MCG+07--16--01 & E/S0 & 07 25 14.4 & +42 00 50.6 & \,\,\,\,\, 9W & \,\,\,\,\, 7S & \,\,\,\,\,\,\,\,\, \,\,\,\,18.5 \\
\hline
\\
\end{tabular}
\end{center}
\end{table}

\section{Summary}
\label{summary}

We used available imaging, photometric and astrometric information (generally from POSS-I, SERC, POSS-II, and SDSS),
as well as deep literature search to determine and characterize
the nature of 39 unconfirmed SNe (from \cite{hakobyan12} database).
Our analysis shows that 21 objects are real SNe, 2 are Galactic stars,
4 are probable SNe, and 12 objects remain unconfirmed.

According to the morphology of the host galaxies,
SNe blue colors, and absolute discovery magnitudes, we tentatively assign a SN type to 4 of them.
In addition, the host galaxy of SN~1976N is identified, and
corrected offsets for SNe~1958E, 1972F, and 1976N are calculated.

\begin{acknowledgement}
\,\ \\
\,\ \\
L.S.A., A.R.P., and A.A.H. acknowledge the hospitality of the Institut d'Astrophysique de Paris (France)
during their stay as visiting scientists supported by the Collaborative Bilateral Research Project of
the State Committee of Science (SCS) of the Republic of Armenia and
the French Centre National de la Recherch\'{e} Scientifique (CNRS).
V.Zh.A. is supported by grant SFRH/BPD/70574/2010 from FCT (Portugal) and would further like to acknowledge support by the ERC under the FP7/EC through Starting \, Grant \, agreement No. 239953.
M.T. is supported by the INAF PRIN 2011 ``Transient Universe''.
This work was made possible in part by a research grant from the Armenian National Science and
Education Fund (ANSEF) based in New York, USA.
This research made use of the Asiago Supernova Catalogue (ASC), which is available at
\texttt{http://web.oapd.inaf.it/supern/cat/}, the website of the Central Bureau for Astronomical Telegrams (CBAT),
available\, at\,\, \texttt{http://www.cbat.eps.harvard.edu/lists/Supernovae.html},\,
the\, NASA/IPAC\, Extragalactic\, Database (NED), which is available at \texttt{http://ned.ipac.caltech.edu/}
and operated by the Jet Propulsion Laboratory, California Institute of Technology,
under contract with the National Aeronautics and Space Administration.
Funding for SDSS-III has been provided by the Alfred P.~Sloan Foundation,
the Participating Institutions, the National Science Foundation, and the US Department of Energy Office of Science.
The SDSS-III web site is \texttt{http://www.sdss3.org/}.
SDSS-III is managed by the Astrophysical Research Consortium for the Participating Institutions of the SDSS-III
Collaboration including the University of Arizona, the Brazilian Participation Group,
Brookhaven National Laboratory, University of Cambridge, Carnegie Mellon University, University of Florida,
the French Participation Group, the German Participation Group, Harvard University,
the Instituto de Astrofisica de Canarias, the Michigan State/Notre Dame/JINA Participation Group,
Johns Hopkins University, Lawrence Berkeley National Laboratory,
Max Planck Institute for Astrophysics, Max Planck Institute for Extraterrestrial Physics,
New Mexico State University, New York University, Ohio State University, Pennsylvania State University,
University of Portsmouth, Princeton University, the Spanish Participation Group, University of Tokyo,
University of Utah, Vanderbilt University, University of Virginia, University of Washington, and Yale University.

\end{acknowledgement}

% BibTeX users please use
% \bibliographystyle{}
% \bibliography{}

\end{document}